\documentclass[twocolumn,showpacs,preprintnumbers,amsmath,amssymb,pre]{revtex4}
\usepackage{graphicx}

\begin{document}

\title{On the influence of noise on chaos in nearly Hamiltonian
systems}
\author{P. V. Elyutin}
\email{pve@shg.phys.msu.su}
\affiliation {Department of Physics,
Moscow State University, Moscow 119992, Russia}

\date{\today}

\begin{abstract}
The simultaneous influence of small damping and white noise on
Hamiltonian systems with chaotic motion is studied on the model of
periodically kicked rotor.  In the region of parameters where
damping alone turns the motion into regular, the level of noise
that can restore the chaos is studied.  This restoration is
created by two mechanisms: by fluctuation induced transfer of the
phase trajectory to domains of local instability, that can be
described by the averaging of the local instability index, and by
destabilization of motion within the islands of stability by
fluctuation induced parametric modulation of the stability matrix,
that can be described by the methods developed in the theory of
Anderson localization in one-dimensional systems.
\end{abstract}
\maketitle

\section{Introduction}

\vspace*{-1mm} From the point of view of chaotic dynamics, the
Hamiltonian systems are marked out by the omnipresence of chaos:
for nearly any Hamiltonian system with not less than one and a
half degrees of freedom (with the exemption of completely
integrable models that are non-robust and therefore exceptionally
rare) the chaotic motion is possible for some initial
conditions.  On the contrary, for the dissipative systems of the
same complexity of the structure chaotic motion on strange
attractors  either could be attained only in limited domains of
the parameter space or is inaccessible at all \cite{1,2}.

Inclusion of the dissipative terms, even arbitrarily small, in the
canonical equations of motion of the Hamiltonian system can change
the character of the motion drastically. In particular, such
addition can banish the chaos: for example, for the autonomous
Hamiltonian systems with added (viscous) damping the only possible
attractors are stable fixed points.  It must be noted that this
abrupt change may be basically formal, resulting from the presence
of the transition to the infinite time limit in the rigorous
definitions of important characteristics of chaotic motion, like
the Lyapunov exponent and correlators of dynamic variables.  In
many experimentally relevant models the ratio of the dissipation
$\gamma$ to the typical frequency of motion $\omega$ may take very
small values.  Thus,  for radiation damping of vibrations of
polyatomic molecules one has $\gamma / \omega \sim 10^{-10}$; the
same order of magnitude of  $\gamma / \omega$ turns out for the
tidal friction of the celestial bodies of the Solar system.  In
these situations the duration of the "transient chaos" phase $T
\sim \gamma^{-1}$ is so long that accurate determination of
characteristics of the chaotic motion can be carried out without
the account of dissipation.

Physically the introduction of dissipation in the equations of
motion is a form of description of the interaction of an isolated
(in the zeroth approximation) system with its environment - a
"heat bath" with practically infinite number of degrees of
freedom, continuous spectrum of eigenfrequencies and internal
dynamics that is independent of the state of the selected
system.  This heat bath may be considered also as a source of
noise - that is, acting on the selected system random forces,
whose statistical characteristics are determined by the properties
of the heat bath alone.  The problem of simultaneous influence of
small dissipation and weak noise on the features of chaotic motion
in the originally Hamiltonian non-autonomous system is the main
concern of this paper.

The studies of the influence of noise on chaotic motion were
pioneered by Lieberman and Lichtenberg \cite{3} just by analysis
of the effect of fluctuations on the Hamiltonian non-autonomous
system. However, the modern paradigm of the domain was formed
later by Crutchfield and Huberman \cite{4,5} who switched the
attention to the exploration of strongly dissipative systems (see
late review \cite{6}).  The influence of noise on the Hamiltonian
systems has been discussed recently in the context of the problem
of decay of metastable chaotic states \cite{7,8}, but in general
the field doesn't seem to be fully investigated.

On the contrary, the influence of small dissipation on the
Hamiltonian chaos is well understood: Afraimovich, Rabinovich, and
Ugodnikov \cite{9} have shown, that with switching on a small
dissipation phase trajectories of stable periodic motions of the
Hamiltonian non-autonomous systems become attractors with regular
motion, and chaos disappears.  With the further increase of
$\gamma$ these attractors may lose their stability; annihilation
of the last one turns the system back into chaotic motion on a
strange attractor, that resembles the chaotic motion of the
original Hamiltonian system.  This pattern needs two
specifications. First, the strange attractor may emerge before
vanishing of the last of regular ones - the system could be
multistable.  This case, mentioned in \cite{9} as "logically
possible", will be met in our model. Secondly, if the Hamiltonian
system has no islands of stability that correspond to periodic
motion, then the transition from Hamiltonian to dissipative chaos
can occur immediately. This case, apparently, will be present in
our model too.

The aforesaid permits to specify the main problem of our paper:
what intensity of noise is necessary to restore the chaos,
repressed by dissipation?

The rest of the text is organized as follows.  In Sec. II the
basic model is introduced.  Sections III and IV treat two
mechanisms of restoration of chaos by noise: fluctuation transfer
to domains of local instability and parametric destabilization of
motion within stability islands.  Sec. V treats the influence of
strong noise on the Lyapunov exponent and correlation functions of
chaotic motion.  Sec. VI contains the summary of results and their
discussion.

\section{THE BASIC MODEL}

We start from the well-known periodically kicked rotor - the
non-autonomous model with the Hamiltonian
\begin{equation}\label{1}
H\left( {I,\theta ,t} \right)={{I^2} \over 2}+K\cos \theta
\sum\limits_{n=-\infty }^\infty  {\delta \left( {t-n} \right)},
\end{equation}
where $I$ and $\theta$ are the dynamic variables (canonically
conjugated momentum - action and coordinate - angle), $K$ is the
control parameter, and $\delta(z)$ is the Dirac delta-function.
The stroboscopic mapping that links values of dynamic variables at
the moments of time $n-0$ and $n+1-0$, preceding two consequent
kicks,
\begin{equation}\label{2}
I'=I+K\sin \theta ,\,\,\,\,\,\,\,\,\,\,\,\theta '=\theta +I',
\end{equation}
is known as the standard, or Chirikov - Taylor, mapping and
thoroughly studied \cite{10}.

The generalization of the model (\ref{1}) that includes
dissipation and noise will be described by the equation of motion
for the angular variable
\begin{equation}\label{3}
\ddot \theta +\gamma \dot \theta -K\sin \theta \sum\limits
_{n=-\infty }^\infty {\delta \left( {t-n} \right)=\xi \left( t
\right)},
\end{equation}
where $\gamma$ is the damping constant. The $\xi(t)$ in the RHS is
the Langevin random force, that is a stationary, distributed by a
Gaussian law, $\delta$-correlated random process (white noise)
with zero mean and the correlator
\begin{equation}\label{4}
\left\langle {\xi \left( t \right)\xi \left( {t+\tau } \right)}
\right\rangle =2\gamma \Theta \delta \left( \tau  \right),
\end{equation}
where $\Theta = k_BT$ is the noise temperature in energy units
(in the system of scales of the model).  The model given by Eqs.
(\ref{3}) and (\ref{4}) has three parameters: $K, \gamma$ and
$\Theta$. We shall restrict ourselves by the domain of small
damping, $\gamma \ll 1$, where the system is nearly Hamiltonian.

In the absence of noise, at $\Theta = 0$, the stroboscopic mapping
for this model is given by equations
\begin{equation}\label{5}
I'=a\left( {I+K\sin \theta } \right),\,\,\,\,\,\,\,\,\,\,\,\theta
'=\theta +b\left( {I+K\sin \theta } \right),
\end{equation}
where \vspace{-2mm}
\begin{equation}\label{6}
a=e^{-\gamma },\,\,\,\,\,\,\,\,\,\,\,b={\gamma ^{-1}({1-e^{-\gamma
}})} .
\end{equation}
The two-parameter mapping given by Eqs. (\ref{5}) is a special
case of the four-parameter Zaslavsky mapping, that has been
introduced in \cite{11} and studied in \cite{12,13,14}.  The main
efforts of these studies were applied to the case $\gamma \gtrsim
1$. Here we describe in brief the properties of our model for the
case of small damping, $\gamma \ll 1$.

At small and moderate values of the control parameter $K$ the
most important attractor of the mapping Eqs. (\ref{5}) is the
fixed point $I=0, \theta = \pi$, that is stable in the range
\begin{equation}\label{7}
K<K_1={{2\left( {1+a} \right)} \over b}\approx 4\left( {1+{7 \over
{12}}\gamma ^2} \right).
\end{equation}
For $K>K_1$ the leading attractor is the symmetric cycle $C^s_2$
of two points that are related by equations $I'=-I,
\theta'=2\pi-\theta$.  It is stable while
\begin{equation}\label{8}
K<K_2={{\pi \left( {1+a} \right)} \over b}\approx 2\pi \left(
{1+{7 \over {12}}\gamma ^2} \right).
\end{equation}
For $K>K_2$ the leading attractors are two asymmetric cycles of
length two $C^{a1}_2$ and $C^{a2}_2$.  The phase coordinates of
their points are related by equations $I'=-I$, $\theta' = \pi +
\theta $. They are stable in the domain
\begin{eqnarray}\label{9}
\nonumber K<K_3={1 \over b}\sqrt {\pi ^2\left( {1+a}
\right)^2+3+a^2}\\
\approx 2\sqrt {\pi ^2+1}\left( {1+0.152\gamma }
\right).
\end{eqnarray}

In general case the system defined by Eqs. (\ref{5}) is
multistable. From the first of these equations it follows that the
strip
\begin{equation}\label{10}
\left| I \right|\le I_+={K \over {\exp  \gamma  -1}}
\end{equation}
is the trap (the absorbing set) of the system: any phase
trajectory that comes within this strip never leaves it. To
determine the comparative roles of basins of attraction of strange
and regular attractors of the model the fraction $f$ of chaotic
trajectories among the set with random initial conditions,
uniformly distributed within the trap, was calculated.  The
results are presented in Fig. 1.
\begin{figure}
\includegraphics[width=0.8\columnwidth]{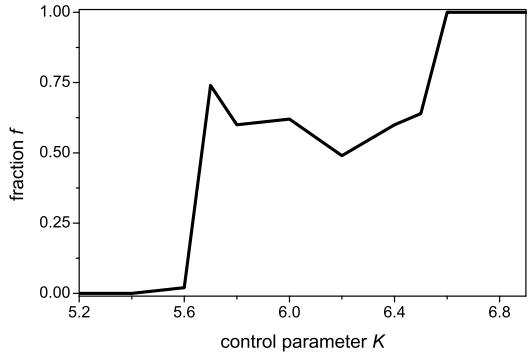}
\caption{\label{fig1} The dependence of the fraction $f$ of the
trap area, covered by the basin of attraction of the strange
attractor of the system Eqs. (\ref{5}), on the control parameter
$K$. The damping value $\gamma = 0.05$.  For each value of $K$ 100
uniformly distributed initial conditions was taken.}
\end{figure}
It can be seen that the strange attractor is born within the
domain of stability of the cycle $C^s_2$, and after the loss of
stability of cycles $C^{a1}_2$ and $C^{a2}_2$, in the range
$K>K_3(\gamma)$, it remains the only apparent attractor of the
system.

For $\Theta > 0$ the stroboscopic mapping for the system Eq.
(\ref{3}) has the form
\begin{eqnarray}\label{11}
\nonumber I'&=&a\left( {I+K\sin \theta } \right)+\upsilon,\\
\theta '&=&\theta +b\left( {I+K\sin \theta } \right)+\varphi,
\end{eqnarray}
where $\upsilon$ and $\varphi$ are the random increments
(fluctuations) of action and angle for the unit interval of time.
Fluctuations $\upsilon$ and $\varphi$ at the moment of time $t$
after the beginning of the motion with definite initial conditions
have the Gaussian distributions with the dispersions
\begin{eqnarray}\label{12}
\nonumber D_\upsilon &=&\Theta \left( {1-e^{-2\gamma t}}
\right),\\
D_\varphi &=&{\Theta  \over {\gamma ^2}}\left( {2\gamma
t-3+4e^{-\gamma t}-e^{-2\gamma t}} \right)
\end{eqnarray}
respectively [15].  In our case $\gamma t \equiv \gamma \ll 1$ and
the expressions Eqs. (\ref{12}) could be replaced by their
asymptotics
\begin{equation}\label{13}
D_\upsilon =2\gamma \Theta
,\,\,\,\,\,\,\,\,\,\,\,\,\,\,\,D_\varphi ={2 \over 3}\gamma
\Theta.
\end{equation}

Fluctuations $\upsilon$ and $\varphi$ are positively correlated;
that has clear physical meaning: positive increment in velocity
(that is numerically equal to the action) at the unit interval of
time leads, most probably, to the positive increment of
coordinate. The correlator $M=\left\langle {\upsilon \varphi }
\right\rangle$ by the moment $t$ after the beginning of motion can
be calculated by the method described in [15]:
\begin{equation}\label{14}
M={\Theta  \over \gamma }\left( {1-e^{-\gamma t}} \right)^2.
\end{equation}
For $\gamma t \equiv \gamma \ll 1$ the asymptotic value of this
correlator is $M=\Theta \gamma$.  The joint distribution of
fluctuations of action and angle has the form
\begin{equation}\label{15}
w\left( {\upsilon ,\varphi } \right)={{\sqrt 3} \over {2\pi \gamma
\Theta }}\exp \left[ {-{1 \over {\gamma \Theta }}\left( {3\varphi
^2-3\varphi \upsilon +\upsilon ^2} \right)} \right].
\end{equation}

In the presence of noise the phase trajectory can reach any point
of the phase space of the system.  However, for finite $\gamma$
the system with overwhelming probability will stay in the strip
with limited action values that is much narrower than the trap
given by Eq. (\ref{10}). For the description of this domain of
concentration of the probability density the term "attractor" will
be used.

\section{THE THRESHOLD OF CHAOS:  TRANSFER TO THE DOMAIN OF
LOCAL INSTABILITY}

The condition of existence of chaos is, by definition, the
positive value of the Lyapunov exponent $\sigma$. Numerical
calculation show, that in our model at moderate values of $K
\lesssim 5.4$, when the motion of the system in the absence of
noise is regular, the Lyapunov exponent increases with $\Theta$
and at some value of $\Theta_0$ passes through zero.

We turn to the theoretical description of the onset of chaos. For
conservative (area-preserving) mappings with strong chaos rather
accurate estimate for the Lyapunov exponent could be obtained by
averaging of the local instability index - the logarithm of the
maximal in absolute value eigenvalue of the stability matrix -
over the domain of chaotic motion \cite{10,16}.  Since our model
is nearly Hamiltonian, we may try to use this approach.

The stability matrix for the mapping given by Eqs. (\ref{11}) is
\begin{equation}\label{16}
A= \left |\begin{matrix}
  \,\,a & aK\cos \theta  \\
  \,\,b & {1+bK}\cos \theta \,\
\end{matrix}\right |
\end{equation}
The local instability index depends only on the angle $\theta$;
for $K<4$
\begin{equation}\label{17}
\sigma \left( \theta  \right)=\ln \left|\, {{S \over 2}+\sqrt
{\left( {{S \over 2}} \right)^2-D}}\, \right|,
\end{equation}
where $S=a+1+bK\cos \theta $ is the trace of $A$ and $D=a$ is its
determinant. For small $\gamma$ almost everywhere in the interval
$\pi / 2 < \theta < 3\pi /2$ the index is negative and constant,
$\sigma (\theta) = -\gamma / 2$, and the motion is locally stable.
Most probably the system stays in this domain, but under the
influence of noise it can sporadically enter the domains of local
instability.  For large enough values of $\Theta$ their
contribution can compensate the weak squeezing of phase
trajectories in the central part of the attractor.

\subsection{Small $K$}

For small values $K \ll 1$ in the absence of damping ($\gamma =
0$) the evolution of the periodically kicked rotor nearly
everywhere in the phase space can be described by the time
averaged (and thus time-independent) Hamiltonian of the system,
that is given by Eq. (\ref{1}):
\begin{equation}\label{18}
\overline {\mathstrut H}\left( {I,\theta } \right)=\overline
{H\left( {I,\theta ,t} \right)}={{I^2} \over 2}+K\cos \theta.
\end{equation}
For an autonomous Hamiltonian system the inclusion of viscous
damping and connection to the Langevin (white noise) heat bath
lead to the canonical distribution of probability in the phase
space
\begin{equation}\label{19}
W\left( {I,\theta } \right)=N \exp \,\,-{{\overline {\mathstrut
H}\left( {I,\theta } \right)} \over \Theta },
\end{equation}
where $N$ is the normalization constant.  For the averaging
$\sigma =\left\langle {\sigma \left( \theta  \right)}
\right\rangle $ one needs to know the angular distribution $W
(\theta)$.  Its normalized form could be found from Eqs.
(\ref{18}) and (\ref{19}):
\begin{equation}\label{20}
W\left( \theta  \right)={1 \over {2\pi I_0\left( {{K
\mathord{\left/ {\vphantom {K \Theta }} \right.
\kern-\nulldelimiterspace} \Theta }} \right)}}\exp \,\,-{K \over
\Theta }\cos \theta,
\end{equation}
where $I_0(z)$ is the zeroth order modified Bessel function of
the first kind. Since nearly all probability density is
concentrated in the stability interval, where $\cos \theta <0$,
the contribution of this domain to the averaged value is
\begin{equation}\label{21}
\sigma _-\approx -{\gamma  \over 2}.
\end{equation}

To calculate the contributions of zones of local instability we
neglect the damping; then we have $\sigma \left( \theta
\right)\approx \sqrt {K\cos \theta }$.  Positive contribution of
two instability strips could be estimated by the integral
\begin{equation}\label{22}
\sigma _+\approx {1 \over {\pi I_0\left( {{K \mathord{\left/
{\vphantom {K \Theta }} \right. \kern-\nulldelimiterspace} \Theta
}} \right)}}\int\limits_{{{3\pi } / 2}}^{2\pi } {\exp \,\,\left(
{-{K \over \Theta }\cos \theta } \right)}\sqrt {K\cos \theta
}\,d\theta.
\end{equation}

The threshold values $\Theta_0$ found by averaging of the local
instability index and by direct numerical calculation are compared
in Fig. 2.

\begin{figure}
\includegraphics[width=0.8\columnwidth]{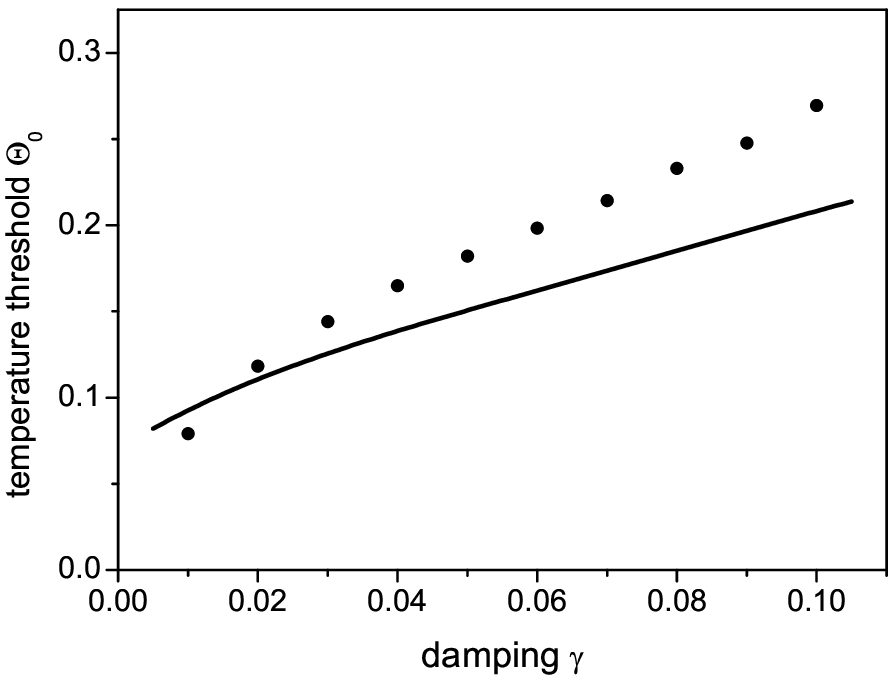}
\caption{\label{fig2} The dependence of the temperature threshold
of chaos $\Theta_0$ on damping $\gamma$ for $K=0.3$. Calculation
by Eqs. (\ref{21}) and (\ref{22}) (line) and numerical calculation
(points).}
\end{figure}

If $K/\Theta \gg 1$, the integral can be calculated analytically:
replacing the Bessel function by its asymptotics for large value
of the argument, and approximating the cosine by the linear
function, we obtain:
\begin{equation}\label{23}
\sigma _+\approx \sqrt {{1 \over {2K}}}\Theta \,\,\exp \,\,-{K
\over \Theta }.
\end{equation}
From the condition $\left\langle \sigma  \right\rangle =\sigma
_-+\sigma _+=0$ the threshold value of temperature is determined
by the root of the equation
\begin{equation}\label{24}
\gamma =\sqrt {{2 \over K}}\Theta_0 \exp \,\,-{K \over \Theta_0 }.
\end{equation}
This expression yields the asymptotic dependence of the
temperature threshold of chaos for small $K$: it has the form
\begin{equation}\label{25}
\Theta _0\approx {K \over {\left| {\,\ln \gamma \,} \right|}}
\end{equation}
and possesses a logarithmic accuracy.

\subsection{Large $K$}

The results of the previous subsection shows that for small $K$
the threshold $\Theta_0$ grows with the increase of the control
parameter $K$, $\Theta_0 \propto K$.  On the other hand, as it
was noted in the Sec. II (see Fig. 1), for $K > K_3$ chaos in the
dissipative system exists (apparently for any initial conditions)
even without any noise. For the reasons of continuity we may
expect that there may exist a range of values of $K$ where the
dependence $\Theta_0(K)$ at the constant damping $\gamma$ is
decreasing. The numerical calculation confirms this suggestion
(see Fig. 3).

\begin{figure}
\includegraphics[width=0.8\columnwidth]{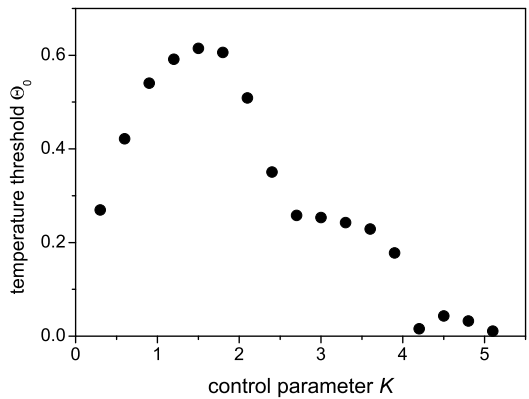}
\caption{\label{fig3} Numerically found dependence of the
temperature threshold of chaos $\Theta_0$ on the control parameter
$K$ for the damping value $\gamma = 0.1$.}
\end{figure}

For the area-preserving mapping Eq. (\ref{2}), reduced to the
basic square $\left( {0\le I<2\pi } \right)\times \left( {0\le
\theta <2\pi } \right)$, for large $K$ the chaotic motion takes
place in the chaotic component that covers the largest part of
the phase space (for $K>2$ the measure of the chaotic component
$\mu (K) > 0.78$).  For the system with damping and noise Eq.
(\ref{11}) we shall retain the name "chaotic component" for the
part of the attractor that includes the points of the chaotic
component of the conservative system, and the complementary part
will be referred to as "islands of stability". Limiting ourselves
to the case $K<4$, we will take into account only one island of
stability that surrounds the stable fixed point $I=0,\,\theta
=\pi $.

Some properties of chaotic motion of the system with damping and
noise in the chaotic component could be described by the
following simple model.  The action variable for one time step
receives the increment $\Delta I\approx K\sin \theta $  with the
averaged square value $\left\langle {\Delta I^2} \right\rangle
\approx {{K^2} \mathord{\left/ {\vphantom {{K^2} 2}} \right.
\kern- \nulldelimiterspace} 2}$.  For the motion in the chaotic
component the correlations of the consequent values of $\theta$
are small \cite{17,18}, and we can depict the evolution of the
system as the motion of the rotor with the damping $\gamma$ under
the influence of some Langevin force, a white noise coming from
the source with an effective temperature $\Theta^{\star}$.  From
Eq. (\ref{13}) we have the estimate $\Theta ^{\star} \approx
K^2/{4 \gamma} \gg 1$.  In this approximation the distribution of
phase density in the chaotic component will become canonical one,
with uniform distribution of angles and the Gaussian distribution
of action,
\begin{equation}\label{26}
W_c\left( {I,\theta } \right)=\sqrt {{\gamma  \over {2\pi
^3K^2}}}\exp \,\left( {- {{2\gamma } \over {K^2}}I^2} \right).
\end{equation}
This expression is applicable for small $\gamma$ and large $K$.

Let's assume that in the island of stability the probability
density  also has the distribution of the canonical form
\begin{equation}\label{27}
W_i\left( {I,\theta } \right)=N\exp \,\left( {-{{\tilde H\left(
{I,\theta } \right)} \over \Theta }} \right),
\end{equation}
where $\tilde H(I, \theta)$ is an effective Hamiltonian (a
function that is constant on the invariant curves of the standard
mapping (2)), and $N$ is the normalization constant. This
assumption is plausible in view of Eq. (\ref{19}); additional
support for it will be obtained in the next section.

If $\gamma$ and $\Theta$ are sufficiently small, then the phase
trajectory can leave the island of stability or return to it only
by passing through the narrow strip of the width $\delta \sim
\sqrt {\gamma \Theta }$ along the border of the island of
stability. The probability $P$ of finding a phase point in the
chaotic component could be found from the balance considerations
by equalizing $W_c$ and $W_i$ on this border. For $P \ll1$ we can
neglect the non-uniformity of the distribution Eq. (\ref{26}) in
action and obtain the estimate
\begin{equation}\label{28}
P=N\sqrt {{{2\pi ^3K^2} \over \gamma }}\exp \left( {-{\Delta \over
\Theta }} \right),
\end{equation}
where $\Delta$ is the value of the effective Hamiltonian ${\tilde
H\left({I,\theta } \right)}$ on the border of the island of
stability.  At present we can not calculate this quantity
analytically, but from its value taken from the numerical
calculations (and depending only on $K$) by Eq.(\ref{28}) we can
find the dependence of $P$ on $\Theta$ and $\gamma$.

In the numerical experiment the basic square $\left( {0\le I<2\pi
} \right)\times \left( {0\le \theta <2\pi } \right)$ has been
separated into $10^4$ cells. A chaotic trajectory of the standard
mapping Eq. (\ref {2}) in this square was calculated for $10^5$
time steps, and all cells, in which the trajectory came at least
once, were marked as the mask of the chaotic component.  Then for
the trajectories of the system with damping and noise
Eq.(\ref{11}) the probability $P$ has been calculated as the
fraction of points of the trajectory whose projections on the
basic square got into one of the cells of the mask.

The results of numerical calculations for the values $K=3$ and
$\gamma = 0.05$ are shown in  Fig. 4.   Fit of the linear
dependence of $\ln P$ on the inverse temperature $\beta = \Theta
^{-1}$ for these points gives values $\Delta = 1.07$ and
$N=0.057$.

\begin{figure}[!ht]
\includegraphics[width=0.8\columnwidth]{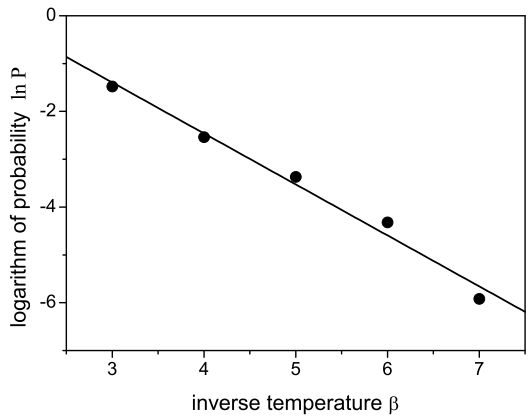}
\caption{\label{fig4} Dependence of the probability $P$ of finding
a trajectory of motion of the system with damping and noise in the
place of the chaotic component of  the standard mapping on the
inverse temperature $\beta = \Theta ^{-1}$ for the values $K=3$
and $\gamma = 0.05$.  Numerical calculation (points) and linear
fit to the points (line).}
\end{figure}

From the assumption that the motion inside the stability island
gives to the Lyapunov exponent the negative contribution
$\sigma_-= -\gamma /2$, and that the positive contribution from
the motion in the chaotic component is $\sigma _+=\sigma \left( K
\right)P$, where $\sigma (K)$ is the Lyapunov exponent value of
the Hamiltonian system, for the temperature threshold of chaos we
obtain the equation
\begin{equation}\label{29}
\Theta _0={{\Delta \left( K \right)} \over {\ln \left[ {\sigma
\left( K \right)N\sqrt {{\mathstrut}8\pi ^3K^2\gamma ^{-3}}}
\right]}}.
\end{equation}

The threshold values $\Theta_0$ found by calculating the
probability of transfer to the chaotic component by this formula
and by direct numerical calculation are compared in  Fig. 5.  From
the Eq. (\ref{29}) it is seen that essentially $\Theta_0$ is
proportional to the "activation energy" $\Delta (K)$; the
dependence on other parameters is only logarithmic.  The general
behaviour of the dependence on  Fig. 3 could be explained by
decrease of the size of the stability island with the increase of
$K$; the dip around $K=4$ reflects the restructurization of the
regular attractor.

\begin{figure}[!ht]
\includegraphics[width=0.8\columnwidth]{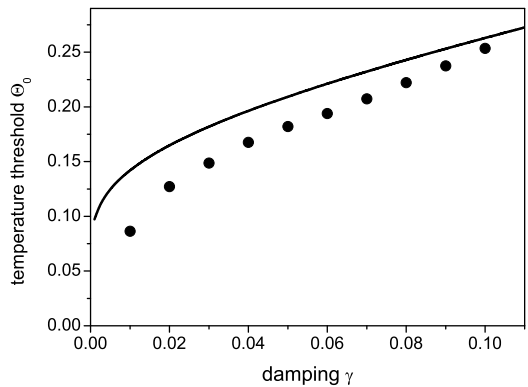}
\caption{\label{fig5} The dependence of the temperature threshold
of chaos $\Theta_0$ on damping $\gamma$ for $K=3$. Calculation by
Eq. (\ref{29}) (line) and numerical calculation (points).}
\end{figure}

\section{THE THRESHOLD OF CHAOS:  PARAMETRIC DESTABILIZATION IN THE
ISLANDS OF STABILITY}

Although the agreement between the theoretical curve and the
numerical points in Fig. 5 is rather convincing, the increase of
discrepancy at very small $\gamma$ is strange: just in this
domain the damping must have especially little influence, and the
picture of transfer between the island of stability and the
chaotic component promises to be asymptotically exact.
Furthermore, this discrepancies could not be neglected from the
quantitative point of view, since for the sharp dependence of
$P(\Theta)$ small variations of $\Theta$ at low temperatures
produce large changes in the positive contribution to the
Lyapunov exponent $\sigma _+=\sigma \left( K \right)P$.  For
example, for $K=3$ and $\gamma = 0.01$ substitution of the
numerically found value $\Theta _0 = 0.086$ in the Eq. (\ref{28})
gives $P=2.6\cdot 10^{-5}$ and $\sigma _+ = 1.8 \cdot 10^{-5}=
3.5 \cdot 10^{-3} (\gamma / 2)$. Thus we must conclude that there
exists another mechanism creating the instability that acts on
the parts of phase trajectories that are localized within the
islands of stability.

In the domain $2<K<4$, in which the island of stability surrounds
the fixed stable point $I=0, \theta = \pi$, we can consider the
dynamic variables $x=\theta - \pi$ and $y=I$ to be small. Then by
substitution $\cos x \approx 1- x^2/2$ the matrix of stability
could be represented in the form
\begin{equation}\label{30}
\tilde A=\left| \begin{matrix}
\,\,a&{-aK+a\eta \,\,}\\
\,\,b&{1-bK+b\eta} \,\,\end{matrix} \right|,
\end{equation}\vspace{2mm}
where $\eta = Kx^2/2$ are small corrections.  The quantities
$\eta$ are fluctuating under the influence of noise and could be
treated as random.  The stochastic modulation of parameters of the
mapping leads to destabilization of the motion. The corresponding
exponent of instability $\sigma _+$ we will calculate in the
conservative approximation, since the contribution to the common
Lyapunov exponent from damping $\sigma _-=-\gamma  /2$ and from
stochastic modulation $\sigma _+$ for small $\gamma$ are additive.

The transformation of variables determined by matrices Eq.
(\ref{30}) at $\gamma =0$ can be reduced to the three-term
recurrent relation for the angular variable:
\begin{equation}\label{31}
x_{n+1}-\left( {2-K-\eta _n} \right)x_n+x_{n-1}=0.
\end{equation}
This expression can be interpreted as an equation for amplitudes
$x_n$ of the stationary wave function in the quantum
one-dimensional tight-binding model with unit non-diagonal matrix
elements (transfer integrals) between adjacent sites, random site
energies $\eta_n$ and the energy eigenvalue $E=2-K$
(one-dimensional chain with diagonal disorder).  The calculation
of the Lyapunov exponent for this system was carried out in the
context of the theory of Anderson localization.  For the case in
which $\eta_n$ are independent random variables with zero mean,
$\left\langle \eta \right\rangle =0$, and small dispersion,
$\left\langle {\eta ^2} \right\rangle \ll 1$, the Lyapunov
exponent has been calculated by Derrida and Gardner \cite{19}.
Correlations between consequent values of $\eta_n$ were taken into
account by Tessieri and Izrailev \cite{20}.  The stochastic
Lyapunov exponent is proportional to the dispersion of fluctuating
parameter $\eta$:
\begin{equation}\label{32}
\sigma _+={{\left\langle {\eta ^2} \right\rangle } \over
{2\left({4K-K^2} \right)}}C\left( \omega  \right).
\end{equation}
The correlation factor $C(\omega)$ in this expression has the form
\begin{equation}\label{33}
C\left( \omega  \right)=1+2\sum\limits_{k=1}^\infty  {b_\eta
\left( k \right)\cos \left( {2\omega k} \right)},
\end{equation}
where $b_{\eta}(k)$ are normalized correlation functions of the
random variable $\eta$, $b_\eta \left( k \right)={{\left\langle
{\eta _i\eta _{i+k}} \right\rangle } \mathord{\left/ {\vphantom
{{\left\langle {\eta _i\eta _{i+k}} \right\rangle } {\left\langle
{\eta ^2} \right\rangle }}} \right. \kern- \nulldelimiterspace}
{\left\langle {\eta ^2} \right\rangle }}$, and
\begin{equation}\label{34}
\omega =\arccos {{2-K} \over 2}
\end{equation}
is the average angle of rotation of a vector by the linearized
standard mapping.  Formula Eq. (\ref{32}) is applicable for the
values of $K$ that are not too close to $K=0$ or $K=4$.  In our
case $\left\langle \eta \right\rangle \ne 0$, but we can include
this value in the parameter $K$; this renormalization will change
its value into $\tilde K=K\left( {1-{{\left\langle {x^2}
\right\rangle } \mathord{\left/ {\vphantom {{\left\langle {x^2}
\right\rangle } 2}} \right. \kern- \nulldelimiterspace} 2}}
\right)$.  In what follows we shall retain the designation $\eta$
for the fluctuating quantity with zero mean, $\eta ={{K\left(
{x^2-\left\langle {x^2} \right\rangle } \right)} \mathord{\left/
{\vphantom {{K\left( {x^2-\left\langle {x^2} \right\rangle }
\right)} 2}} \right. \kern- \nulldelimiterspace} 2}$. To use the
expression Eq. (\ref{32}) we have to determine statistical
characteristics of the variable $\eta$: its dispersion and
correlation function. It must be noted at once that for
calculation of these quantities the account of damping is
essential.

\subsection{Invariant density in the island of stability}

Since for low temperatures $\Theta$ the phase trajectory nearly
all the time is located in the vicinity of the stable fixed point,
for finding the invariant distribution of the probability density
$W(x,y)$ we may use the linearized mapping $ \hat L$,
\begin{equation}\label{35}
x'=by+\left( {1-bK} \right)x,\,\,\,\,\,\,y'=ay-aKx.
\end{equation}
The motion of the system on a unit time interval can be separated
into two stages: the first is the evolution under the mapping Eq.
(\ref{35}), the second is addition of the fluctuation increments
(cf. Eq. (\ref{11})).  The probability of coming in the vicinity
of the point $(u,v)$ after the first stage is proportional to the
value of the invariant density in the vicinity of its prototype
$\hat L^{-1}\left( {u,v} \right)$.  The influence of noise can be
described by the convolution of the obtained distribution with the
distribution of fluctuation increments $w(\upsilon, \varphi)$ Eq.
(\ref{15}).  Thus for the invariant density we obtain the
following integral equation:
\begin{widetext}

\begin{equation}\label{36}
W\left( {x,y} \right)=a^{-1} {\displaystyle {\int\!\!\! \int}}
{du\,dv} \,W\left( {u-{\displaystyle {b \over a}}v,
{\displaystyle {Ku}+{{1-bK} \over a}}v}\right)
{\displaystyle{{\sqrt 3} \over {2\pi \gamma \Theta }}}\exp
\left\{ {\displaystyle{-{1 \over {\gamma \Theta }}}\left[
{3\left( {u-x} \right)^2-3\left( {u-x} \right)\left( {v-y}
\right)+\left( {v-y} \right)^2} \right]} \right\}.
\end{equation}

\end{widetext}

We will look for its solution in the form of a canonical
distribution with the effective Hamiltonian, that is bilinear in
action $y$ and angle $x$:
\begin{equation}\label{37}
W\left( {x,y} \right)=\exp \,-{1 \over \Theta }\left(
{Ax^2+Bxy+Cy^2} \right).
\end{equation}
After substitution of this expression in Eq. (\ref{36}),
integration and equivalizing the coefficients at the identical
powers of dynamic variables, in the lowest order in $\gamma$ we
obtain the parameters of the function Eq. (\ref{37}):
\begin{equation}\label{38}
A={3K \over {6-K}},\,\,\,\,\,\,\,B=-{3K \over
{6-K}},\,\,\,\,\,\,\,C={3 \over {6- K}}.
\end{equation}
Now we can calculate the moments of dynamic variables, e.g.
\begin{equation}\label{39}
\left\langle {x^2} \right\rangle ={{12-2K} \over {12K-3K^2}}\Theta
,\,\,\,\,\,\,\,\,\,\,\,\left\langle {xy} \right\rangle ={{6-K}
\over {12-3K}}\Theta,
\end{equation}
and the dispersion of the fluctuating quantity,
\begin{equation}\label{40}
\left\langle {\eta ^2} \right\rangle ={1 \over 2}\left( {{{12-2K}
\over {12- 3K}}} \right)^2\Theta ^2,
\end{equation}
that enters in the RHS of Eq. (\ref{32}).

\subsection {Correlation function}

With the known form of the invariant density, the correlation
function of the angular variable $x$ could be calculated by the
direct integration.  For the linearized mapping Eq. (\ref{35}) the
values of $B_x\left( n \right)=\left\langle {x_ix_{i+n}}
\right\rangle $ could be expressed through two moments,
$\left\langle {x^2} \right\rangle $ and $\left\langle {xy}
\right\rangle $.  For example,
\begin{eqnarray}\label{41}
\nonumber &\,\,\,\,B_x\left( 0 \right)=\left\langle {x^2}
\right\rangle \,\,\,\,\,\,\,B_x\left( 1 \right)=\left( {1-bK}
\right)\left\langle {x^2} \right\rangle +b\left\langle
{xy} \right\rangle ,\cr \\
\nonumber  &B_x\left( 2 \right)=\left( {1-bK\left( {2+a}
\right)+b^2K^2}
\right)\left\langle {x^2} \right\rangle \\
&+b\left( {a-bK+1} \right)\left\langle {xy{\mathstrut}}
\right\rangle .
\end{eqnarray}
For small $\gamma$ the normalized correlation function $b_x\left(
n \right)={{B_x\left( n \right)} \mathord{\left/ {\vphantom
{{B_x\left( n \right)} {B_x\left( 0 \right)}}} \right.
\kern-\nulldelimiterspace} {B_x\left( 0 \right)}}$ can be
expressed in the form
\begin{equation}\label{42}
b_x\left( n \right)=\cos \left( {\tilde \omega n} \right)\exp
\left( {-{\gamma \over 2}n} \right).
\end{equation}
Here the tilde over $\omega$ reminds that the renormalized value
of $\tilde K$ must be used in calculations.  This formula is
compared to the numerical calculations in Fig. 6.

\begin{figure}[!ht]
\includegraphics[width=0.8\columnwidth]{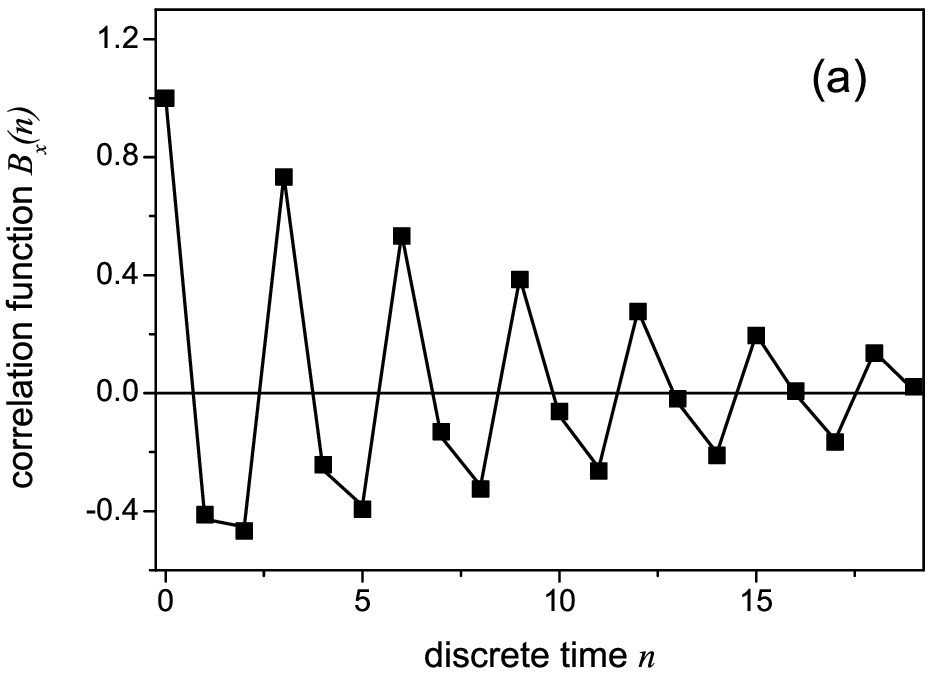}
\end{figure}

\begin{figure}[!ht]
\includegraphics[width=0.8\columnwidth]{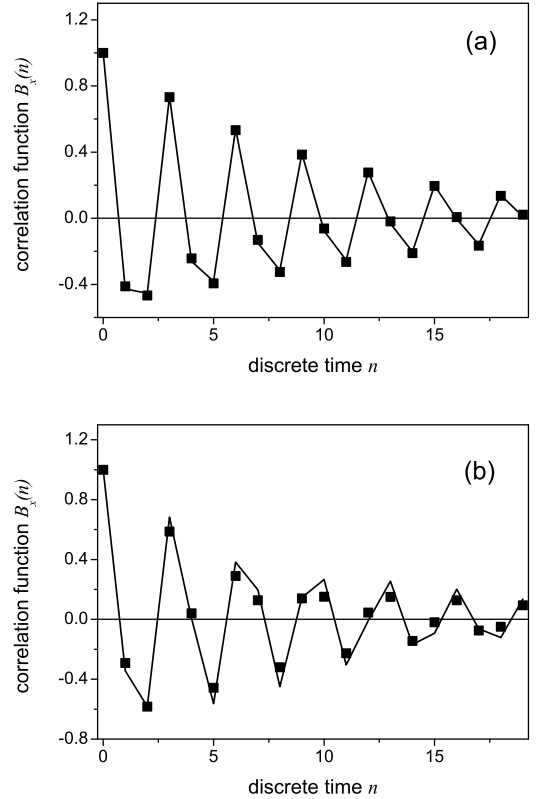}
\caption{\label{fig6} Normalized correlation function $b_x(n)$ of
the angular variable for the values of parameters $K=3$ and
$\gamma = 0.2$. (a) For $\Theta = 0.05$, (b) for $\Theta = 0.2$.
Calculation by Eq. (\ref{42}) (line) and numerical calculation
(points).}
\end{figure}

The normalized correlation function of the fluctuating variable
$\eta$ can be calculated in a similar way:
\begin{equation}\label{43}
b_\eta \left( n \right)=\cos ^2\left( {\tilde \omega n}
\right)\exp \left( {-\gamma n} \right).
\end{equation}
With this expression one can calculate the correlation factor
$C(\omega)$ (see Eq. (\ref{33})).  For small $\gamma$ it is given
by the expression $C\left( \omega \right)\approx \left( {2\gamma
} \right)^{-1}$.  With it from the condition $\sigma_-+\sigma_+=0$
follows the estimate of the temperature threshold of chaos
\begin{equation}\label{44}
\Theta _0=2\sqrt {4K-K^2}{{12-3K} \over {12-2K}}\gamma.
\end{equation}

This formula is too crude for the practical application: it gives
only the estimate of $\Theta_0$ from below.  Here is the reason
for this limitation: the expressions Eqs. (\ref{42}) and
(\ref{43}) for the correlation functions are valid only for small
temperatures, $\Theta \lesssim \gamma$. For larger values the
nonlinear terms that are present in the exact mapping Eq.
(\ref{11}) change the frequency of oscillations of the correlation
function $b_{\eta}(n)$ (see Fig. 6(b)); by this they spoil the
resonance with the cosine factor under the summation sign in Eq.
(\ref{33}) and considerably decrease the value of $C(\omega)$,
down to the value about 2-3 on the threshold of chaos.  The
numerical calculation shows that around $\Theta_0$ the temperature
dependence of the Lyapunov exponent is accurately described by
the formula
\begin{equation}\label{45}
\sigma =-{\gamma  \over 2}+\kappa \Theta ^2
\end{equation}
with a coefficient $\kappa$ about unity.  From Eq. (\ref{45}) the
simple approximation for the temperature threshold follows:
\begin{equation}\label{46}
\Theta _0\approx c\sqrt \gamma,
\end{equation}
where constant $c$ is about unity.  The fit of the law Eq.
(\ref{46}) to the points in Fig. 5 gives $c \approx 0.88$.

\section{STRONG NOISE}

In this section we will look at the effects of noise with
temperature much higher than the chaos threshold $\Theta_0$.

\subsection {The Lyapunov exponent}

With the increase of the noise temperature the Lyapunov exponent
increases monotonously and tends to some finite limit
$\sigma_{\infty}$ (see Fig. 7).

\begin{figure}
\includegraphics[width=0.8\columnwidth]{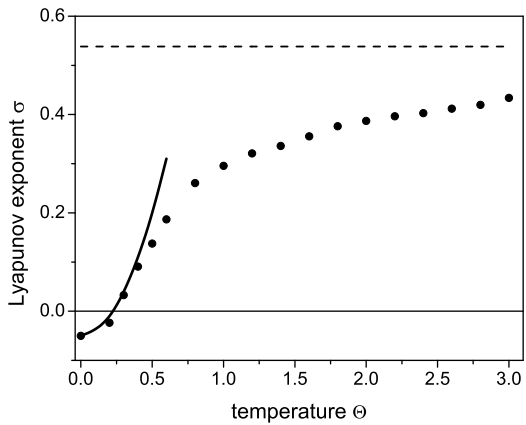}
\caption{\label{fig7} The dependence of the Lyapunov exponent
$\sigma$ on the noise temperature $\Theta$ for $K=3$ and $\gamma =
0.1$ obtained from numerical calculation (points). Parabola in the
left part (solid line) - calculation by Eq. (\ref{45}) with
$\kappa = 1$. Horizontal dashed line marks the limiting value
$\sigma _\infty =0.538$.}
\end{figure}

Since with the increase of noise the typical values of increments
of the angle ($\varphi$) and of the action ($\upsilon$) grow,
$\varphi \sim \upsilon \sim \sqrt {{\mathstrut} \gamma \Theta }$
(see Eq. (\ref{13})), for $\Theta \gg \gamma^{-1}$ all
correlations of dynamical variables vanish. Therefore the limiting
value $\sigma_{\infty}$ is equal to the Lyapunov exponent of the
infinite product of the matrices Eq. (\ref{17}) with uncorrelated
values of $\theta$, uniformely distributed in the interval $0 \le
\theta < 2\pi$. For small $K$ the value of $\sigma_{\infty}$ can
be calculated from the result of \cite{19} for the localization
length at the band edge, namely:
\begin{equation}\label{47}
\sigma _\infty =-{\gamma  \over 2}+0.229K^{{2 \mathord{\left/
{\vphantom {2 3}} \right. \kern-\nulldelimiterspace} 3}}.
\end{equation}
The dependence given by this equation is in good agreement with
the numerical data up to the value of $K \approx 1$ (see Fig. 8).

\begin{figure}
\includegraphics[width=0.8\columnwidth]{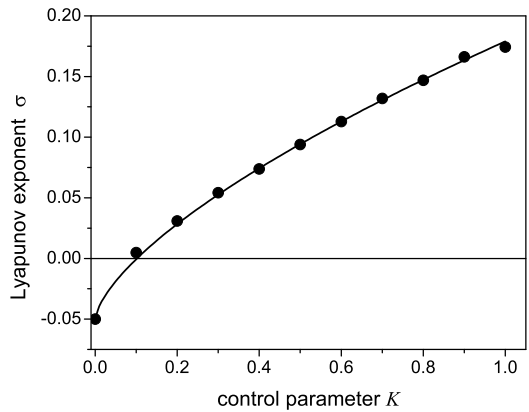}
\caption{\label{fig8} The dependence of the limiting value of the
Lyapunov exponent $\sigma _{\infty}$ on the control parameter $K$
for $\gamma =0.1$.  Calculation by Eq. (\ref{47}) (line) and
numerical calculation (points).}
\end{figure}

It may be noted that from Eq. (\ref{47}) it follows that for a
given value of damping $\gamma$ there exist a range of values of
the control parameter $K<3.23\,\gamma ^{{3 \mathord{\left/
{\vphantom {3 2}} \right. \kern-\nulldelimiterspace} 2}}$ in
which chaos could not be reached for any intensity of the noise.
For large values $K \gtrsim 3$ the limit $\sigma_{\infty}$ does
not differ noticeably from the Lyapunov exponent $\sigma (K)$ of
the original Hamiltonian system.

\subsection{The angular correlations}

In the theory of the standard mapping it is customary to study
the angular correlations through the properties of the variable
$s = \sin \theta$ \cite{17,18}.  From the symmetry considerations
it has zero mean: $\left\langle s \right\rangle =0$.

Let's consider the correlation of two consequent values of this
variable:
\begin{equation}\label{48}
B_s\left( 1 \right)=\left\langle {\sin \theta \,\,\sin \theta '}
\right\rangle.
\end{equation}
When the invariant density $W(I,\theta)$ is known, the calculation
of the correlation Eq. (\ref{48}) is reduced to the two-fold
integration.  For small $K$ the distribution can be taken as the
canonical one with the averaged Hamiltonian Eq. (\ref{18}). With
approximating the angular distribution by the Gaussian function,
for small damping $\gamma$ we have the expression
\begin{equation}\label{49}
B_s\left( 1 \right)={1 \over 2}\left( {1-\exp \left( {-{{2\Theta
} \over K}} \right)} \right)\exp \left( {-{\Theta  \over 2}}
\right).
\end{equation}
It is compared to the numerical data in Fig. 9(a).

\begin{figure}
\includegraphics[width=0.8\columnwidth]{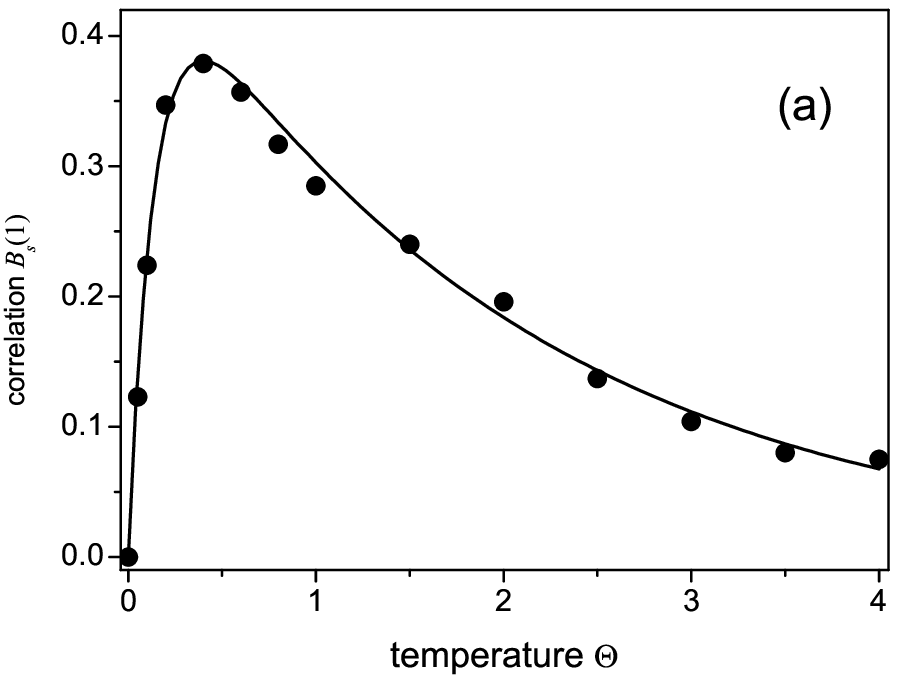}
\end{figure}

\begin{figure}
\includegraphics[width=0.8\columnwidth]{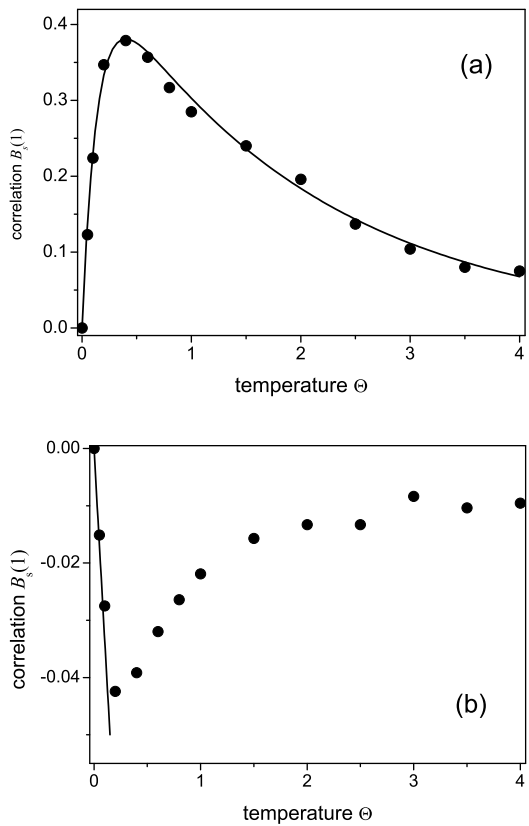}
\caption{\label{fig9} Dependence of the correlation function
$B_s(1)$ of the angular variable $s=\sin \theta$ on the noise
temperature $\Theta$. (a) For $K=0.3$, $\gamma = 0.01$.
Calculation by Eq. (\ref{49}) and numerical calculation (points).
(b) For $K=3$, $\gamma = 0.05$.  Calculation by Eq. (\ref{50})
(line) and numerical calculation (points).}
\end{figure}

For large values of $K \ge 2$ and small temperatures the value of
$B_s(1)$ could be calculated from the distribution Eq. (\ref{37}):
\begin{equation}\label{50}
B_s\left( 1 \right)=\left( {1-K} \right)\left\langle {x^2}
\right\rangle +\left\langle {{\mathstrut xy}} \right\rangle
={{12-8K+K^2} \over {12K-3K^2}}\Theta.
\end{equation}
It is compared to the numerical data in Fig. 9(b). The numerical
calculation shows in this case the decrease of $B_s(1)$ with
$\Theta$ for sufficiently high temperatures.  The reason for it
is qualitatively clear.  We have two competing contributions to
$B_s(1)$: a negative one from the island of stability and a
smaller positive one from the motion in the chaotic component.
The increase of the temperature leads to the increase of the
probability $P$ of the motion within the chaotic component, that
eventually suppresses the negative contribution. The way of
accurate calculation of $B_s(1)$ for high temperatures at present
is not known.

Thence both for small and large values of the control parameter
$K$ the increase of the noise temperature (from zero) induces
first the increase of the correlation of the consequent values of
the angular variable $s=\sin \theta$ up to some maximal value,
and then its decrease.

\section{CONCLUSION}

Above we have studied the model of periodically kicked rotor with
added damping and white noise.  We expect that some of the
established features and relations are typical and will hold at
least qualitatively for many representative non-autonomous
Hamiltonian systems with chaotic motion.  With this in view, in
this Section we summarize main results of this paper in a
generalized way. Since the growth of the control parameter of the
standard mapping $K$ increases both the amount (given by the
invariant measure $\mu$ of the chaotic component) and the
intensity (given by Lyapunov exponent $\sigma$) of chaos we will
refer to $K$ as to the strength of chaos.  In what follows (as
well as everywhere above) the Lyapunov exponent denotes the
largest of the characteristic exponents of the stroboscopic
mapping of the system that can take negative as well as positive
values (one have to keep in view that for systems - flows with
finite phase velocity there is always one zero characteristic
Lyapunov exponent that corresponds to evolution of the
infinitesimal displacement along the phase trajectory).

\textbf{1.} If chaos in a Hamiltonian system is suppressed by
addition of small dissipation, then addition of white noise can,
as a rule, restore the chaotic motion.   The exception is found
only for very weak chaos, when the system remains regular at
arbitrarily intense noise (see Sec. V.A).  The increase of the
noise intensity raises the Lyapunov exponent. In wide context this
fact is not quite trivial, since there is an example of the system
for which noise diminishes the (positive) Lyapunov exponent,
eventually turning it negative \cite{21}.

\textbf{2.}  The temperature threshold of chaos depends on the
strength of chaos of the Hamiltonian system in a non-monotonous
way (see Fig. 3).  For weak chaos it increases with the strength
(in our model -- by linear law, see Eq. (\ref{25})), since the
effective "potential well" corresponding to the island of
stability becomes deeper, whereas for strong chaos the threshold
decreases due to the shrinking of the islands of stability.

\textbf{3.}  There are two essentially different mechanisms of the
chaotization of motion by noise.  The first one is the
fluctuational transfer of the motion from the stability island to
the locally unstable regions of the phase space; its contribution
to the Lyapunov exponent depends on the noise temperature by the
"activation law", $\sigma_+ \propto \exp (-\Delta/\Theta)$ (see
Eqs. (\ref{23}) and (\ref{28}) and Fig. 4). The second is the
parametric destabilization inside the islands of stability created
by small fluctuations of nonlinear terms of the stroboscopic
mapping; its contribution to the Lyapunov exponent depends on the
noise temperature by the power law, $\sigma_+ \propto \Theta ^2$
(see Eqs. (\ref{32}) and (\ref{40}) and Fig. 7). Any one of these
mechanisms could be dominating, depending on the combination of
parameters.

\textbf{4.} Around the threshold of chaos the motion of the
system with damping and noise differs strongly from the chaotic
motion of the original Hamiltonian system.  It is concentrated
mainly within the islands of stability with only rare excursions
to the domain of the phase space occupied by the chaotic
component of the prototype.  In this aspect the restoration of
chaos "by noise" differs radically from the restoration of chaos
"by damping" \cite{9}.  The similarity to the original motion
could be reached in the domain of strong chaos and high noise
temperatures $\Theta \gg 1$ (see Fig. 7).

Lastly it must be noted that the existence of the range of
parameters where the transition from Hamiltonian to dissipative
chaos is immediate (see Fig. 1) may be specific for the studied
model of periodically kicked rotor.  In this range the influence
of noise on chaos has qualitative peculiarities: e.g. the increase
of noise could reduce the Lyapunov exponent (for $K=7$ and $\gamma
= 0.1$ at $\Theta =0$ $\sigma = 1.260(3)$, and at $\Theta = 100$
$\sigma = 1.228(3)$).  The scenario of the immediate transition
and related problems for noisy systems may deserve a special
study.

\section*{ACKNOWLEDGMENTS}

This research was supported by the "Russian Scientific Schools"
program (grant \# NSh - 1909.2003.2).

\end{document}